\input{aipcheck}

\documentclass[
    ,final            
  ]
  {aipproc}

\layoutstyle{6x9}

\begin{document}

\title{ATLAS RPC QA results at INFN Lecce}


\author{M. Bianco}{
  address={\em Dipartimento di Fisica - via Arnesano 73100, Lecce - Italy}
  ,altaddress={\em INFN - via Arnesano 73100, Lecce - Italy} 
}
\author{I. Borjanovic}{
  address={\em Dipartimento di Fisica - via Arnesano 73100, Lecce - Italy}
}
\author{G. Cataldi}{
  address={\em INFN - via Arnesano 73100, Lecce - Italy}
}
\author{A. Cazzato}{
  address={\em Dipartimento di Fisica - via Arnesano 73100, Lecce - Italy}
  ,altaddress={\em INFN - via Arnesano 73100, Lecce - Italy} 
}
\author{G. Chiodini}{
  address={\em INFN - via Arnesano 73100, Lecce - Italy}
}
\author{M.R. Coluccia}{
  address={\em Dipartimento di Fisica - via Arnesano 73100, Lecce - Italy}
  ,altaddress={\em INFN - via Arnesano 73100, Lecce - Italy} 
}
\author{P. Creti}{
  address={\em INFN - via Arnesano 73100, Lecce - Italy}
}
\author{E. Gorini}{
  address={\em Dipartimento di Fisica - via Arnesano 73100, Lecce - Italy}
  ,altaddress={\em INFN - via Arnesano 73100, Lecce - Italy} 
}
\author{F. Grancagnolo}{
  address={\em INFN - via Arnesano 73100, Lecce - Italy}
}
\author{R. Perrino}{
  address={\em INFN - via Arnesano 73100, Lecce - Italy}
}
\author{M. Primavera}{
  address={\em INFN - via Arnesano 73100, Lecce - Italy}
}
\author{S. Spagnolo}{
  address={\em Dipartimento di Fisica - via Arnesano 73100, Lecce - Italy}
  ,altaddress={\em INFN - via Arnesano 73100, Lecce - Italy} 
}
\author{G. Tassielli,}{
  address={\em Dipartimento di Fisica - via Arnesano 73100, Lecce - Italy}
  ,altaddress={\em INFN - via Arnesano 73100, Lecce - Italy} 
}
\author{A. Ventura}{
  address={\em INFN - via Arnesano 73100, Lecce - Italy}
}

\begin{abstract}
 The main results of the quality assurance tests
 performed on the Resistive Plate Chamber used
 by the ATLAS experiment at LHC as muon trigger chambers
 are reported and discussed.
 These are dark current, gas volume tomography, gas tightness,
 efficiency, and noise rate.   
\end{abstract}

\maketitle


\section{Introduction}
Resistive Plate Chamber RPC \cite{Santonico} will be used as 
the muon trigger detector, in the barrel region of the ATLAS experiment at LHC \cite{Atlas}.
A total number of 1116 RPC modules will be installed, 
for a total surface area of about 3800 $m^{2}$.
The extreme difficulty in accessing the
ATLAS detectors, after installation is complete, imposes a high standard quality assurance
for these modules.
For this purpose three cosmic ray teststands have been built at INFN Napoli\cite{Alviggi}, 
Lecce \cite{pic04Boston}, and Roma 2. Here, we report on the results from the Lecce site. 

\section{ATLAS RPC modules}
In the ATLAS muon spectrometer a large variety of sizes and shapes of stations and, 
hence, RPC counters is foreseen. 
All counters, however, share the standard internal structure, described here in the following. 

A unit consists of a doublet of RPC detector layers enclosed in rigid lateral profiles
and two support panels. A thin honeycomb-paper panel with aluminum skin separates 
the two layers realizing two independent Faraday cages. In figure [1] a cartoon of a 
RPC layer and a read-out panel are shown.
A RPC layer contains a $\rm 2 \ mm$ thick active gas layer inside a planar bakelite 
gas volume ($\approx 10^{10}\Omega cm$) externally painted with graphite.  
The gas volume is surrounded by two pick-up strip panels segmented in two orthogonal
views and separated by insulating plastic foils.
The gas volume high voltage electrode is connected directly to the power supply, but the
other one is connected to ground by a shunt resistor, in such a way that the dark current
can be monitored.

The RPC dimension can be as large as 4.8 m in one direction and 1.1 m in the 
other. Depending on the module size, a bakelite gas layer can consist of
a single gas volume or of two adjacent gas volumes, and, most of the times, a
read-out strip plane consists of two adjacent strip panels, instead of one. 
This results in modules which have 2 or 4 gas volumes and 4 or 8 read-out strip panels. 

\begin{figure}
  \includegraphics[height=.2\textheight]{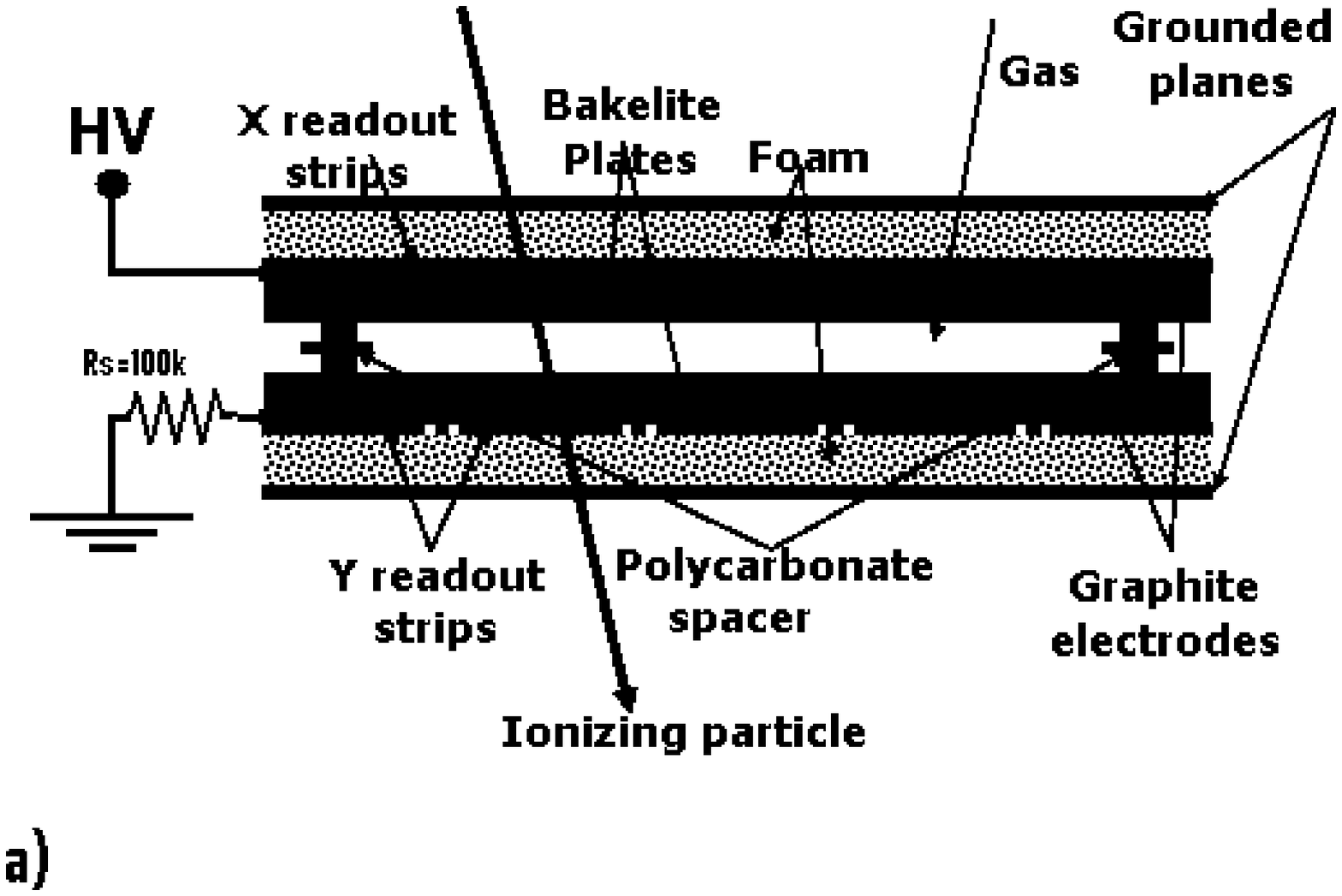}
  \includegraphics[height=.2\textheight]{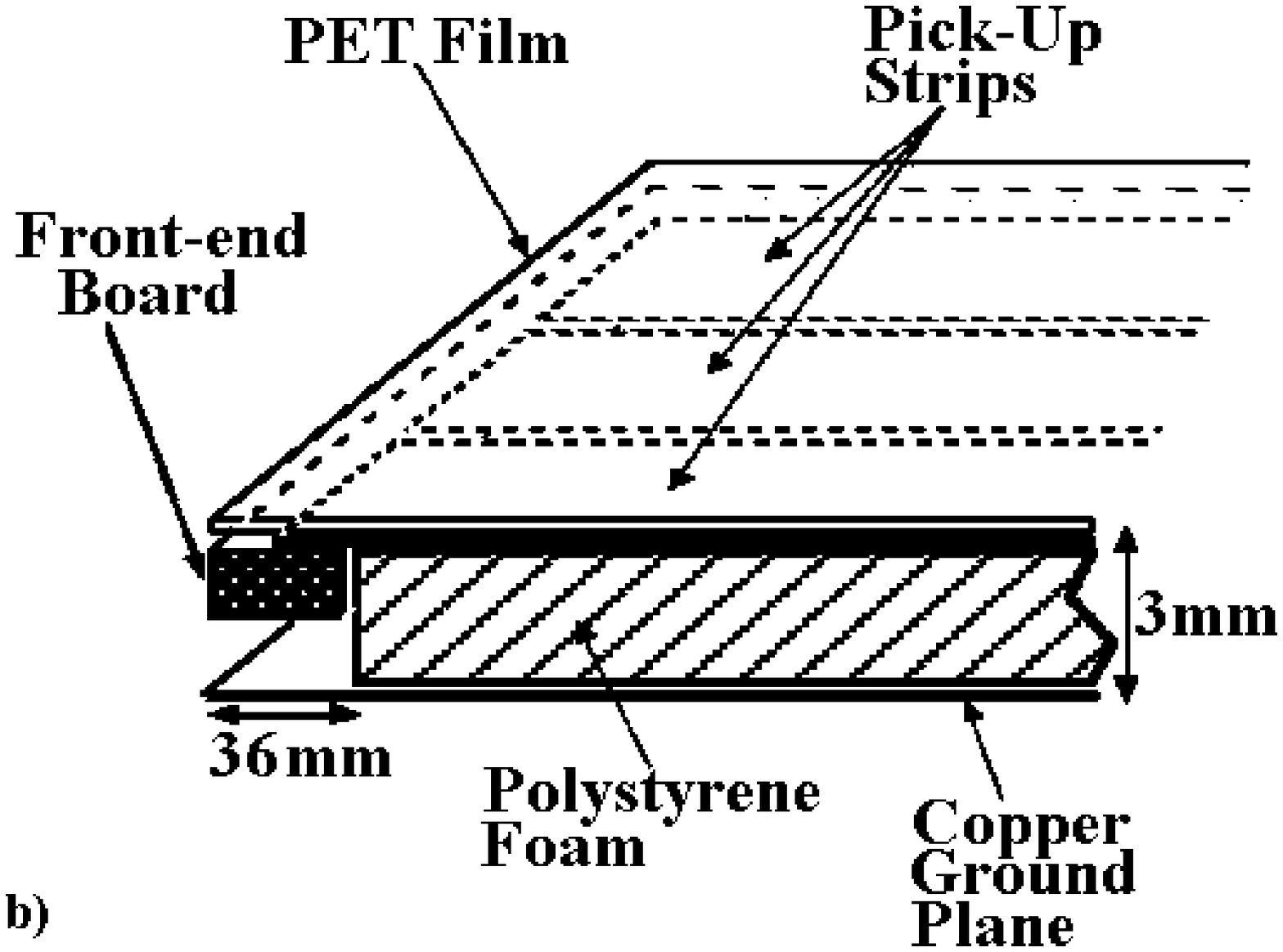}
  \caption{Drawing of ATLAS RPC layer (a) and read-out strip panel (b).}
\end{figure}

\section{Quality assurance procedure}
 The quality control of the RPC chamber is accomplished with a series of 
 accurate measurements and tests intended to verify the correctness of the assembly and 
 detector performance. They represent the very first full chamber characterization and allow 
 to extract statistical information useful to monitor the assembly line and 
 give useful feed-back for improvements.
 
 The quality control procedure consists of main certification tests and 
 subsidiary control tests. The main tests regard leakage current versus high 
 voltage curves, chamber efficiency and noise versus high voltage and front-end 
 voltage threshold, and, finally, chamber 2D tomography. Instead, subsidiary 
 control tests regard pulse test, gas volume leak test, front-end current 
 absorption, and gas volume leakage current temporal drift. 
  
\section{Quality assurance results}
\subsection{Gas volume tests}
The radiation reliance and aging properties of the ATLAS RPC has been proven 
and carefully investigated at X5 gamma irradiation facility at CERN \cite{X5}. 
Nevertheless, the quality control of the gas volumes is crucial. 
The dark current versus high voltage curve is measured during the chamber conditioning
and at the end of the tests. 
Gas volumes showing a dark current with a large ohmic or exponential part are rejected. 
In addition the dark current of each gas volume is monitored continuously
to look for anomalous drift or current glitches 
(maybe due to local discharge or defective high voltage connectors).
In figure [2] on the left the gas volume dark current distribution of ATLAS RPC tested in Lecce
is shown at nominal conditions. 
On the same figure, on the right, is plotted  
the distribution of the difference between the high voltage power supply current 
and the gas volume current.
From this plot is possible to infer a high level of electrical insulation 
between gas volumes and other parts, such as mechanical supports 
and read-out strip panels ($>100 G\Omega$) .
  
\begin{figure}
\includegraphics[height=.28\textheight]{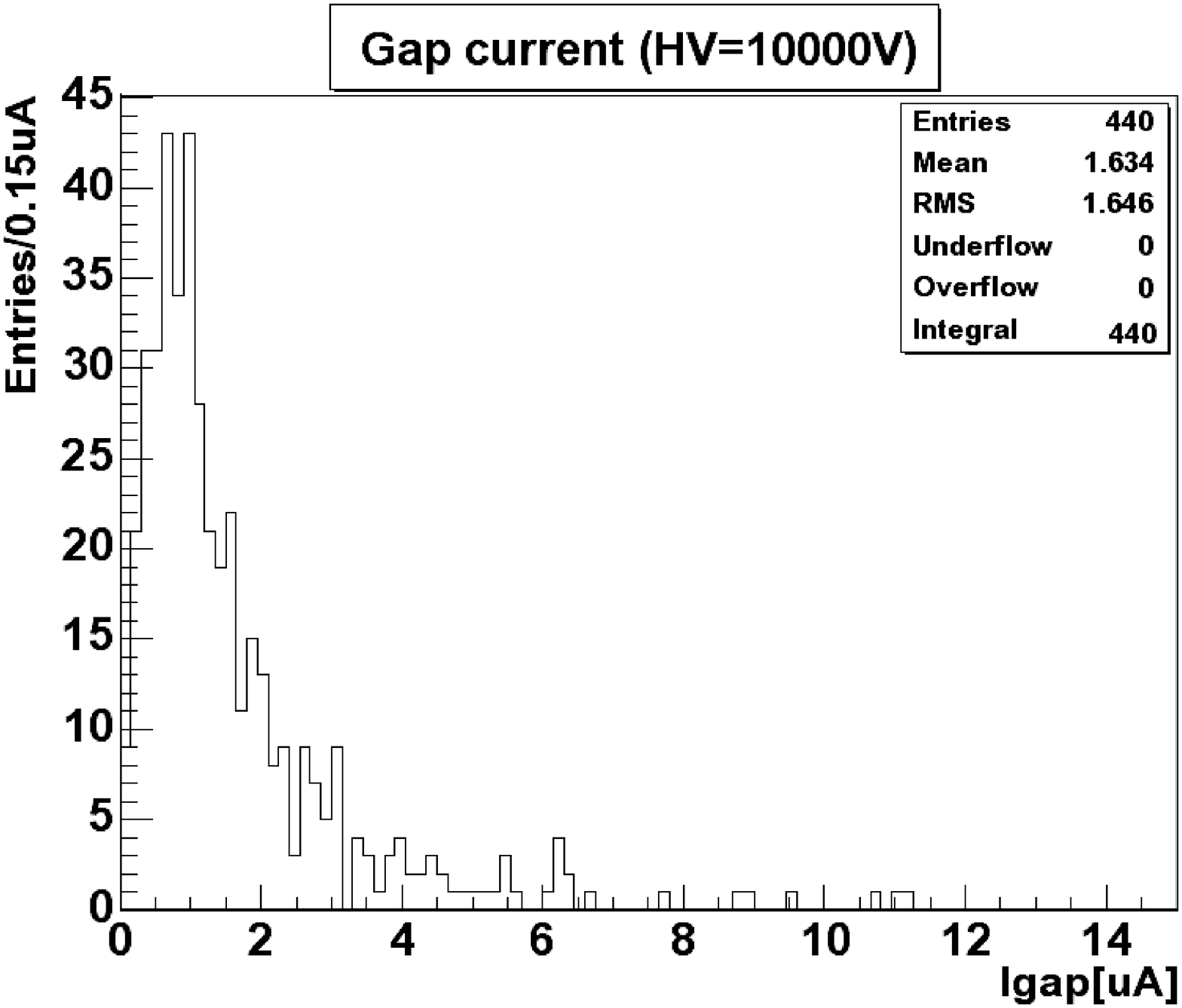}
\includegraphics[height=.28\textheight]{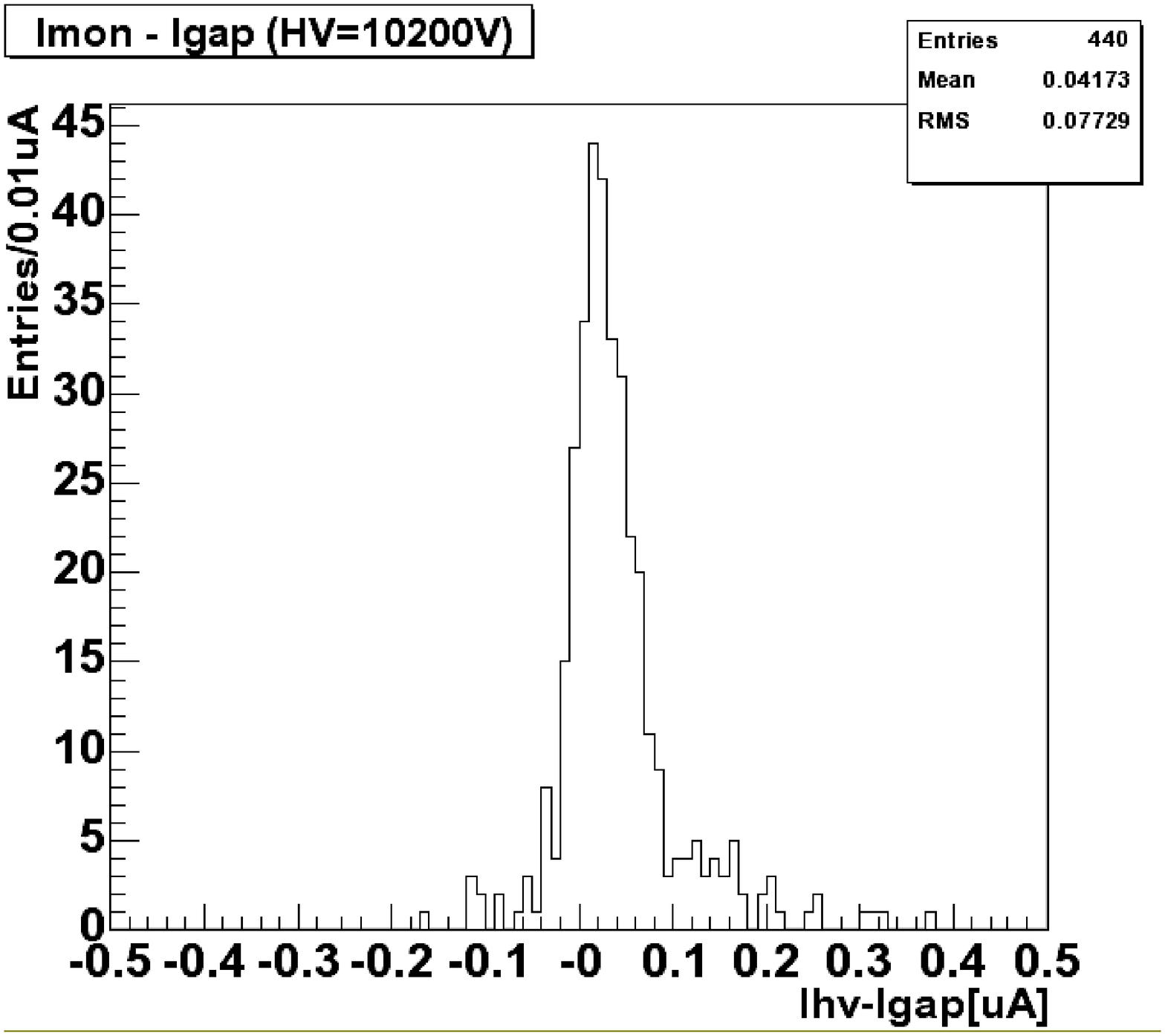}
  \caption{Gas volume dark current distribution (on the left) and power supply current 
  and dark current difference distribution (on the right).
  The data are taken from the ATLAS RPC tested in Lecce with a high voltage of 10kV.}
\end{figure}

The 2D efficiency of each gas volume is measured with high accuracy and this is the
most time consuming test which takes about 24 hours of running. 
The muon cosmic ray tomography is a very powerful tool to discover gas volume assembly
defects (due for example to glue spots, badly glued spacers, and inner 
surface contaminations) which can
compromise the aging performance of the unit. 
A noise rate 2D map is also taken, in order to exclude the presence
of hot spots, which can become degenerative in the long term period.
A gas volume tomography of a special unit is reported in figure [3]. This particular
gas volume has a cut-out in order to leave room for the barrel toroid feet. Other special
counters have gas volumes with cut-out for the alignment laser rays of the MDT precision muon chambers.  
Figure [3] shows that the high quality of the gas volumes is also achieved when 
shapes more complicated than the rectangular one are realized.
  
\begin{figure}
\includegraphics[height=.2\textheight]{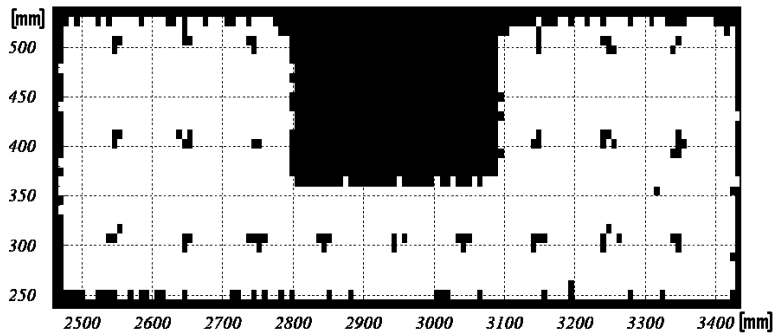}
  \caption{Example of tomography plot of a small size gas volume with cut-out.
 The dark regular spot array are due to the spacers dead area.}
\end{figure}
Gas tightness test, which is performed independently for each layer, consists in 
monitoring the differential pressure for about 2 hours, after 
imposing about 3 mbar of over-pressure and closing the gas inlet and outlet.
With our instrumental sensitivity of about $10^{-1}$ mbar 
we can reject modules with a gas leak larger than about $10^{-4}mbar\frac{l}{s}$.

\subsection{Read-out strip panel tests}
The read-out strip panel efficiency 
(given by the convolution of gas volume efficiency and 
strip panel electronic efficiency) is measured as a 
function of the applied high voltage and front-end voltage threshold.
This allows to establish the detector 
working point and the parameter range where the 
chamber performs well.    
In figure [4] the distribution of the read-out strip 
panel efficiency (dead channels included) and   
the average cluster sizes (number of adjacent firing strips) are shown. 
The RPC counters have in average an efficiency greater than 97\%, 
as expected by the detector active area coverage and required 
by the trigger design.

\begin{figure}
\includegraphics[height=.28\textheight]{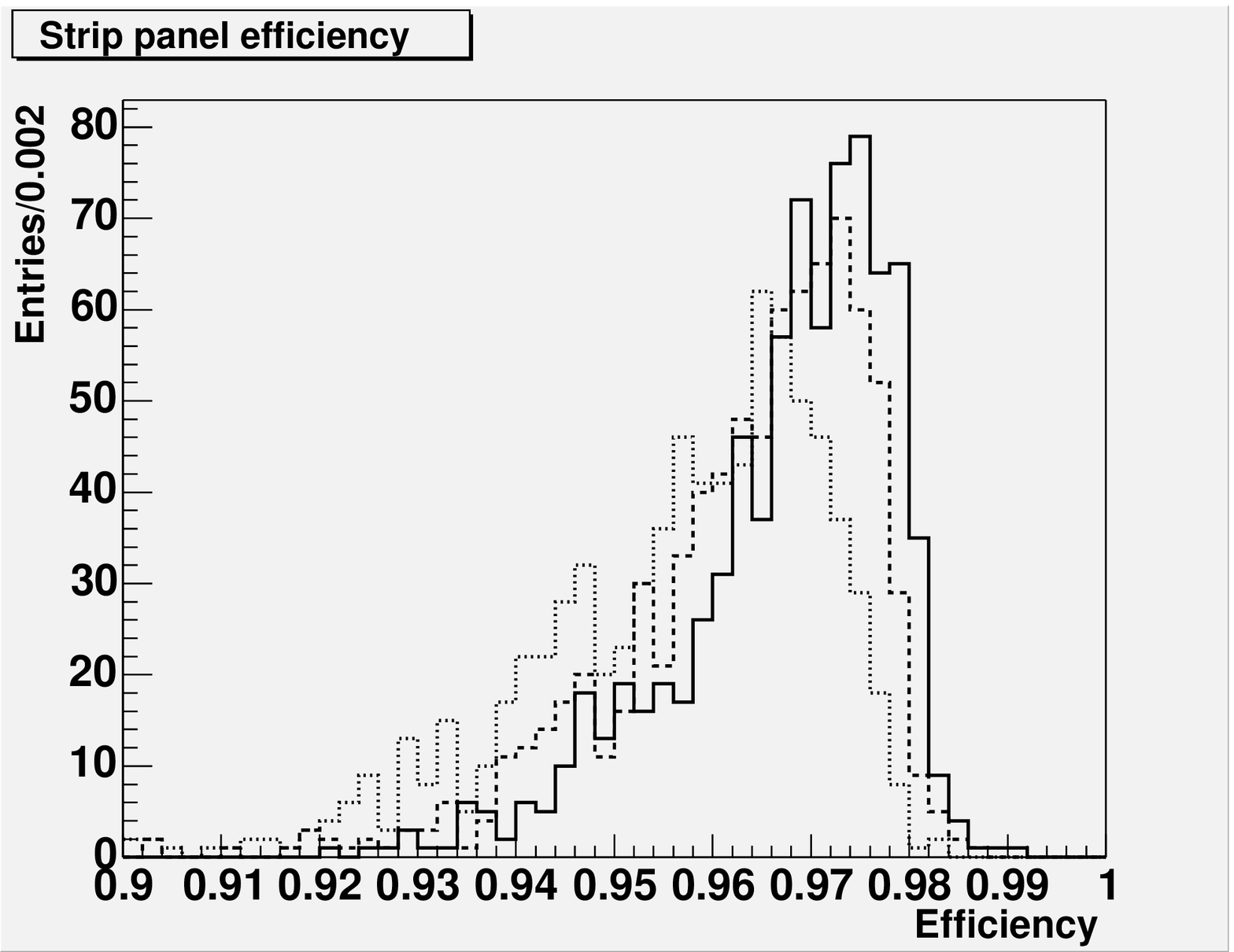}
\includegraphics[height=.28\textheight]{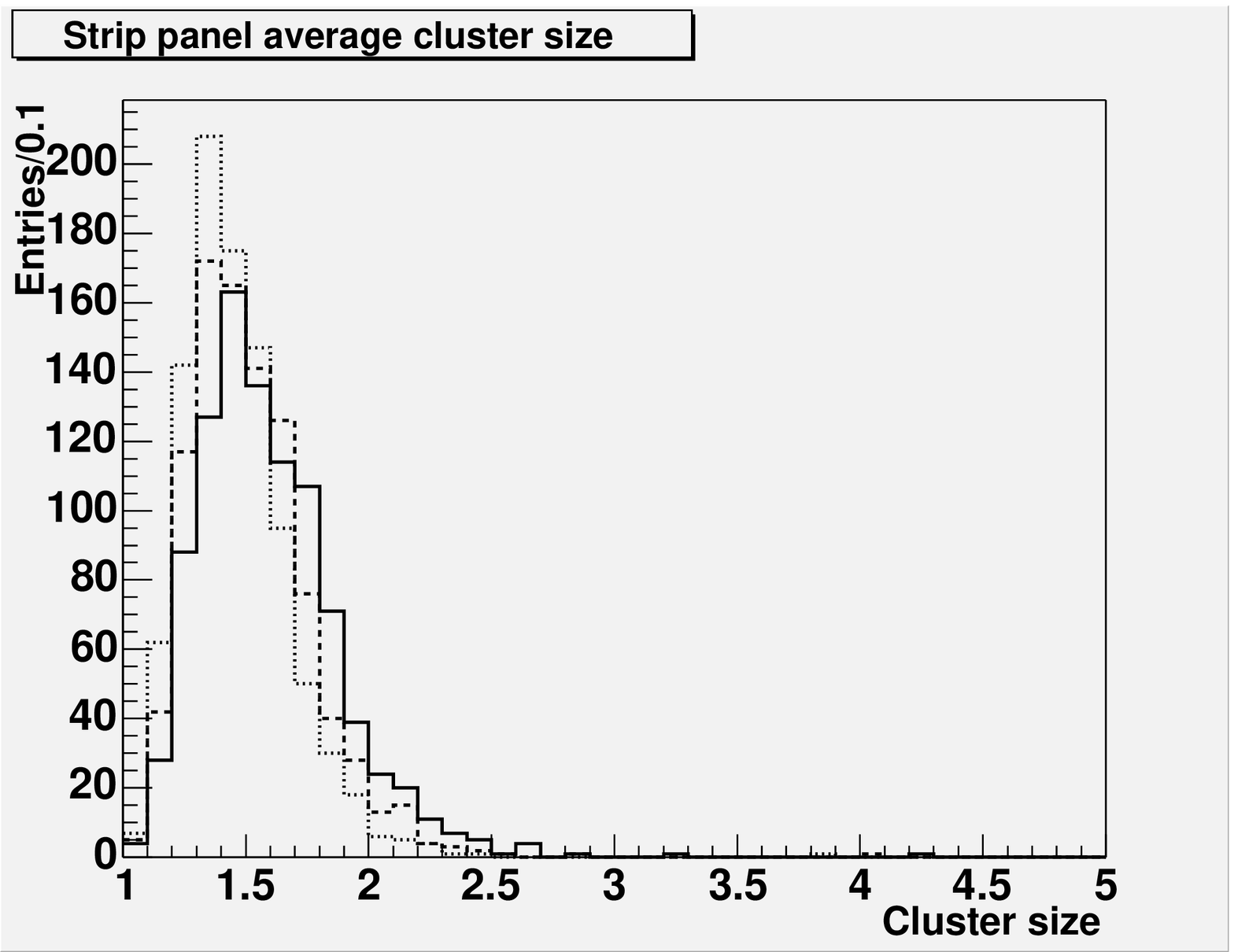}
  \caption{Total efficiency distribution (on the left) and average cluster size distribution (on the right) 
  for different high voltage values of 10kV (...), 10.1kV (.-.), and (-) 10.2 kV.
  The data are taken from the ATLAS RPC tested in Lecce 
  with an equivalent input voltage threshold of about 0.5 mV.}
\end{figure}

The read-out strip panel noise rate is also measured at different high voltage and voltage threshold.
In figure [5] the corresponding distribution is shown on the left.
The average noise rate is less than 1 $\frac{Hz}{cm^{2}}$, which is an order of magnitude smaller than the
expected counting rate due to the photon and neutron background in the ATLAS experiment at 
the designed LHC luminosity.  
In the same figure on the right, we reported the distribution of the ratio between the gas volume dark current 
(with the ohmic component subtracted) and the total strip panel noise rate. By assuming that the
exponential part of the dark current is due to noise counts, we can estimate 
an average saturated avalanche charge of about 15 pC. 
 
\begin{figure}
\includegraphics[height=.26\textheight]{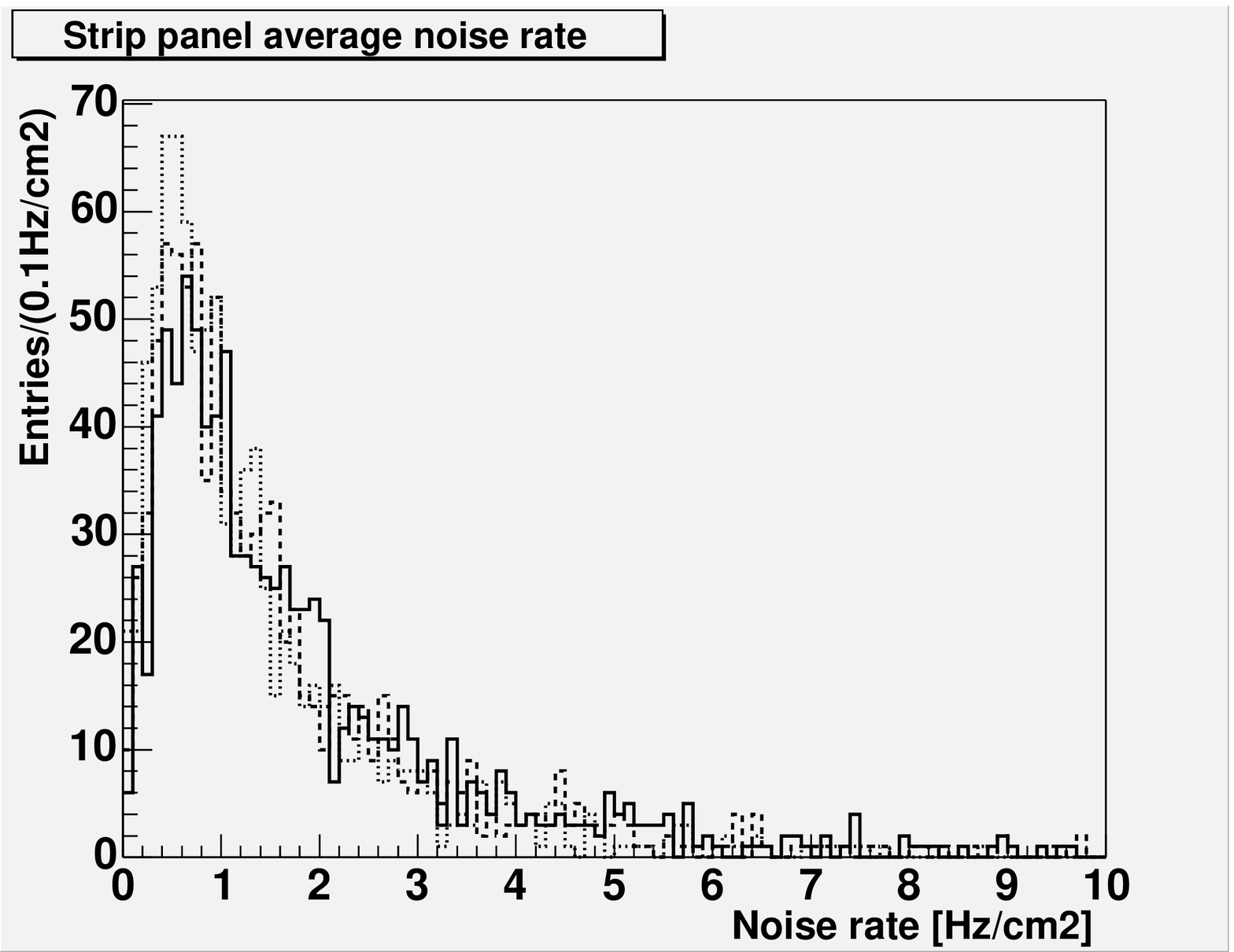}
\includegraphics[height=.26\textheight]{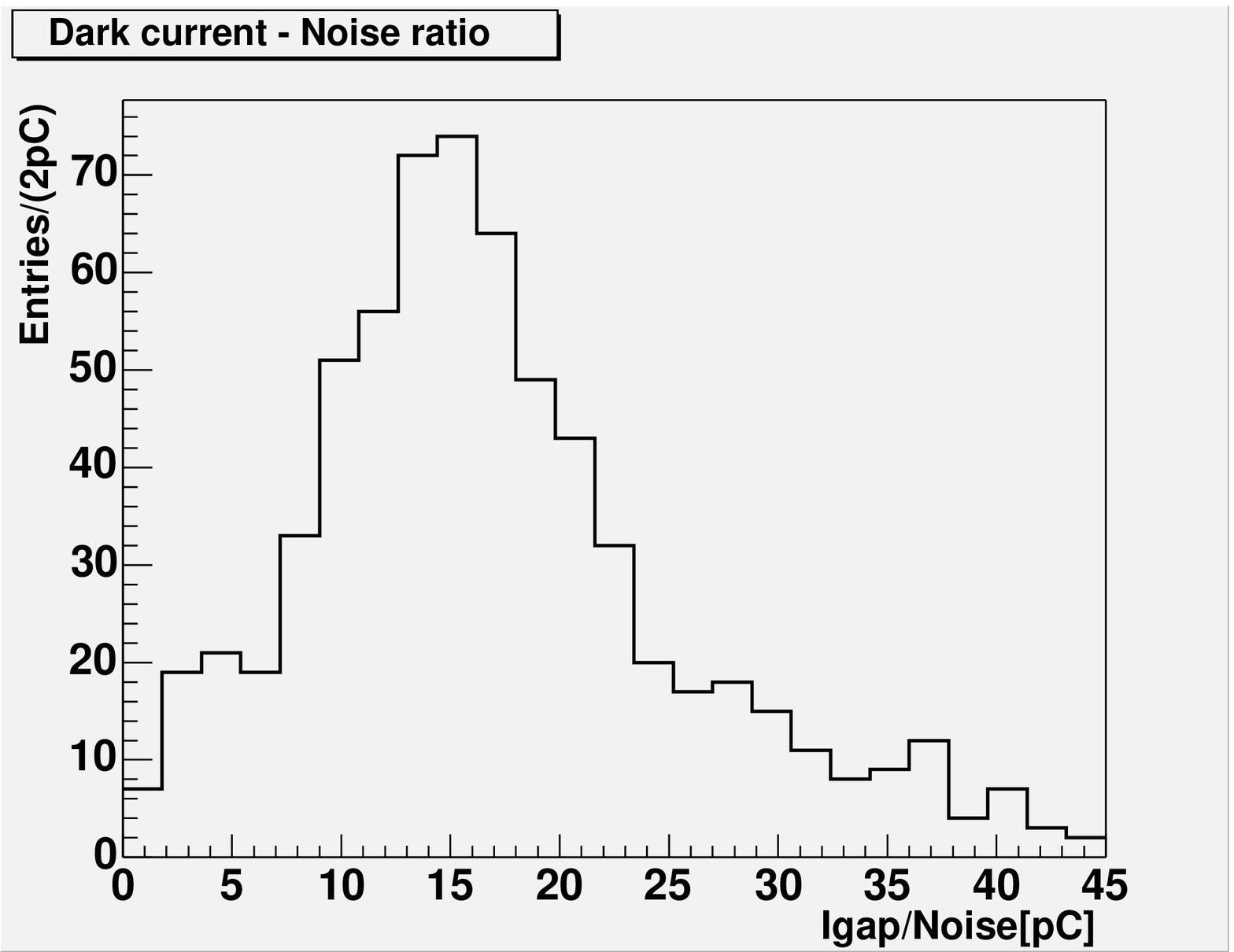}
  \caption{Noise rate distribution (on the left) for different high voltage values of 
  10kV (...), 10.1kV (.-.), and (-) 10.2 kV. Current-to-Noise ratio distribution 
  at a high voltage value of 10kV (on the right).
  The data are taken from the ATLAS RPC tested in Lecce 
  with an equivalent input voltage threshold of about 0.5 mV.}
\end{figure}

\subsection{Production components yield }
Table [1] reports the number of counters which failed the quality assurance test and 
the defective component responsible for the rejection.   
A rejection rate of about 6\% is due to the gas volumes, 
which are discarded, 14\% to the read-out panels, which are repaired, 
and, finally, 6\% to assembly errors.   
Taking into account the number of components inside a module (we tested about 2000 strip
panels and 1000 gas volumes), we have a yield of
98.5\% for gas volume and 98.3\% for the strip panel (without considering the 
factory single component pre-selection).

\begin{table}
\begin{tabular}{lllllll}
\hline
\tablehead{1}{l}{b}{Component failure}
& \tablehead{1}{l}{b}{Number of modules out of 272}
& \tablehead{1}{l}{b}{Fraction of modules}\\
\hline
Readout strip panel & 37 & 14\% \\
Gas volume spot & 9 & 3.5\%\\
Gas volume high dark current & 7 & 2.6\% \\
Gas leak & 9 & 3.5\%\\
HV connector & 2 & 0.9\%\\
Gas volume shunt resistor & 3 & 1.3\%\\

\hline
\end{tabular}
\caption{RPC modules failure rate observed during QA tests in Lecce.}
\label{tab:a}
\end{table}

\section{Conclusions}

The cosmic rays teststand for the ATLAS RPCs in Lecce is routinely 
in operation since July 2004, and up to now, july 2005, 
it has certified about 270 RPC modules,
corresponding to about 25\% of the ATLAS Muon Spectrometer.
The test results and their statistical distributions show
that the RPC properties are stable and uniform, satisfying the ATLAS experiment requirements.
The end of the Quality Assurance Tests is foreseen for the end of October 2005.

\begin{theacknowledgments}
We thank R. Assiro, E.M. Fasanelli, G. Fiore, R. Gerardi, A. Miccoli, 
S. Podladkin and F. Ricciardi for the generous technical assistance 
they provide us in the construction 
and operational phase of the cosmic ray test-stand.
\end{theacknowledgments}

\end{document}